\documentclass{webofc}
\usepackage[varg]{txfonts}   % Web of Conferences font
\usepackage{subcaption}
\usepackage{wrapfig}
\usepackage{tabulary}

\begin{document}
\title{The Scalable Systems Laboratory: a Platform for Software Innovation for HEP}
%
% subtitle is optional

\author{\firstname{Robert} \lastname{Gardner}\inst{1}\fnsep\thanks{\email{rwg@uchicago.edu}}
        \firstname{Lincoln} \lastname{Bryant}\inst{1} \and
        \firstname{Mark} \lastname{Neubauer}\inst{2} \and
        \firstname{Frank} \lastname{Wuerthwein}\inst{3} \and
        \firstname{Judith} \lastname{Stephen}\inst{1} \and
        \firstname{Andrew} \lastname{Chien}\inst{1} 
        % etc.
}

\institute{ Enrico Fermi Institute, University of Chicago, Chicago, IL, USA 
\and
            University of Illinois, Champaign-Urbana, IL, USA 
\and    
            University of California San Diego, La Jolla, CA, USA 
          }
         
\abstract{%
  The Scalable Systems Laboratory (SSL), part of the IRIS-HEP Software Institute, provides Institute participants and HEP software developers generally with a means to transition 
  their R\&D from conceptual toys to testbeds to production-scale prototypes. The SSL enables tooling, infrastructure, and services supporting innovation of novel analysis and data architectures, development of software elements and tool-chains, reproducible functional and scalability testing of service components, and foundational systems R\&D for accelerated services developed by the Institute. The SSL is constructed with a core team having expertise in scale testing and deployment of services across a wide range of cyberinfrastructure. The core team embeds and partners with other areas in the Institute, and with LHC and other HEP development and operations teams as appropriate, to define investigations and required service deployment patterns. We describe the approach and experiences with early application deployments, including analysis platforms and intelligent data delivery systems.
}
\maketitle
\section{Introduction}
\label{intro}
The Institute for Research and Innovation in Software for High Energy Physics (IRIS-HEP)~\cite{IRISHEP} was established 
to meet the software and computing challenges of the HL-LHC.
The Institute is addressing key elements of the
``Roadmap for HEP Software and Computing R\&D for the 2020s''~\cite{CWPDOC}.
IRIS-HEP is one outcome of international and U.S.\ HEP
community planning processes; these were driven in part
by the NSF-funded S2I2-HEP Institute Conceptualization Project~\cite{S2I2HEP}.
Three R\&D areas are of primary interest in IRIS-HEP: 
(1)~development of innovative algorithms for data reconstruction and
triggering (IA); (2)~development of highly performant analysis systems
that reduce `time-to-insight' and maximize the HL-LHC physics
potential (AS); and (3)~development of data organization, management and access (DOMA)
systems for the community's upcoming Exabyte era. 

Each of these R\&D areas requires
resources and service environments for `in-context' development and innovation.  
For this purpose IRIS-HEP created a Scalable Systems Laboratory (SSL). 
The SSL provides the Institute and the HL-LHC experiments with a means to transition 
R\&D to production-scale prototypes. 
The area supports innovation of novel analysis systems and data services, 
functional and scalability testing of service 
components, and foundational systems R\&D for accelerated services developed by the Institute.

\section{SSL Organization and Scope}
\begin{figure}[th]
\centering
\includegraphics[width=0.9\textwidth,height=0.9\textheight,keepaspectratio]{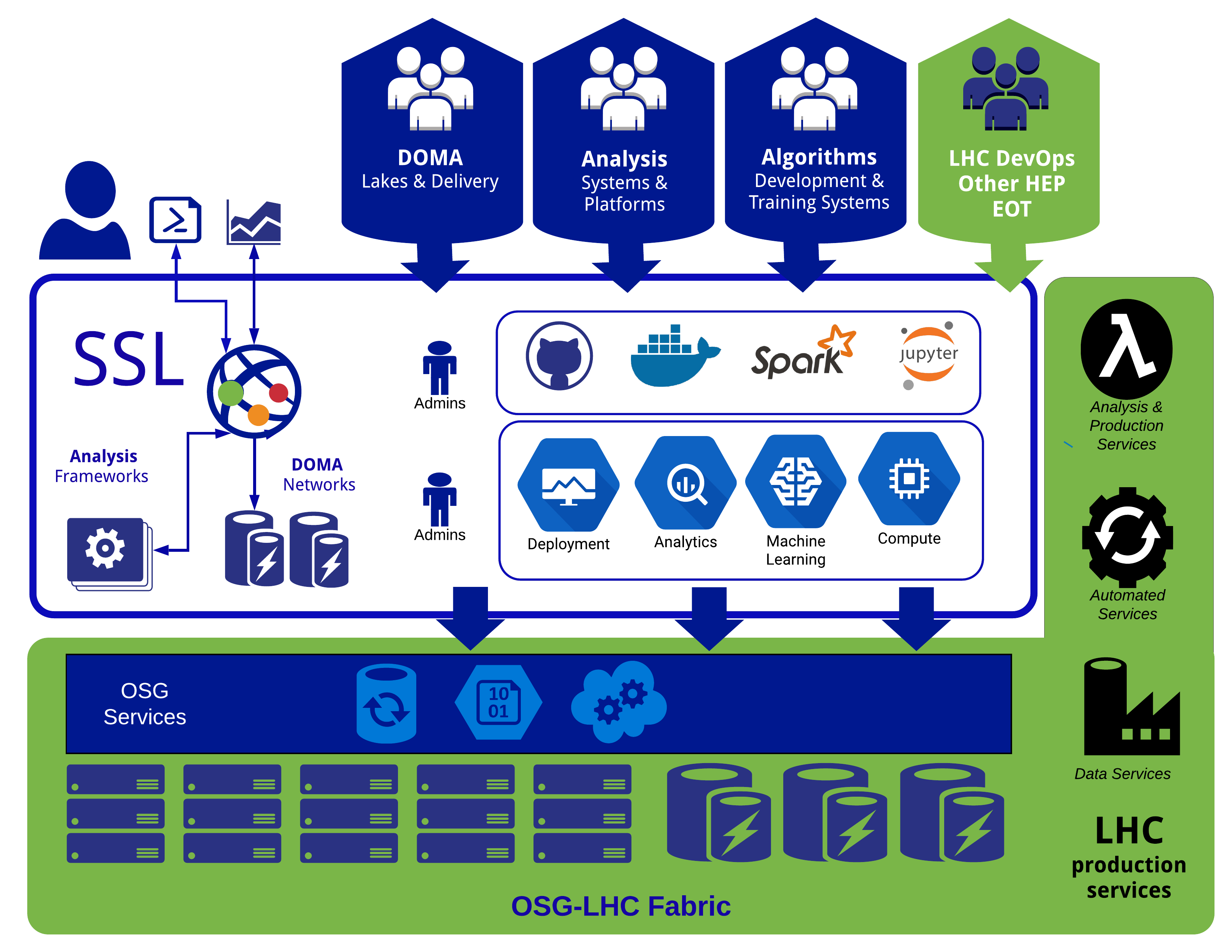}
\caption{A logical depiction of the IRIS-HEP Scalable Systems Laboratory in relation to the R\&D areas of IRIS-HEP, the software and computing teams of the LHC experiments, and the production infrastructure as provided (in this context) by the Open Science Grid (OSG-LHC).  The flexible Kubernetes infrastructure of the SSL resulted from an Institute blueprint workshop to understand
needs for scalable analysis platforms.}
\label{fig:ssl}       % Give a unique label
%\vspace*{-15pt}
\end{figure}

The SSL (schematically shown in Figure~\ref{fig:ssl}) is constructed to 
have a core team with expertise in scale 
testing and deploying services across a wide range of cyberinfrastructure.  
This core team embeds and partners with other areas in the Institute to 
define investigations, design concrete tests, deploy the needed services and dependent infrastructure, 
execute and evaluate the results.  
The team embeds with the relevant area for each test, 
dynamically growing and allowing it to draw upon a wider pool of effort to accomplish its goals.

Scalable platforms created by the SSL area are not intended ultimately to be used as large scale-production
infrastructures, but are to result in \emph{declarative patterns} that can be re-used across many facilities. 
The idea is that partnerships (e-infrastructure projects, research computing facilities at
labs and universities) would be able to `dynamically' join a distributed cyberinfrastructure. 
These might come from a number of resource types: (1) existing U.S. LHC 
computing resources (including personnel and 
hardware platforms), (2) NSF-funded R\&D 
programs such as SLATE~\cite{slatensfweb,SLATE:gw17}, (3) the OSG, (4) NSF supercomputers, and 
(5) institutional resources affiliated with the Institute.
In this context
we will incorporate innovations in automation, service orchestration, and configuration management
in building suitable DevOps environments for the software innovation teams. We have demonstrated 
this approach previously within the community~\cite{SCALEGWMS, SCALEGWMS2, htcondor:scale}.

%\begin{figure}[htbp]
\begin{wrapfigure}{r}{0.4\textwidth}
\begin{center}
\vspace{-.2in}
\includegraphics[width=0.4\textwidth]{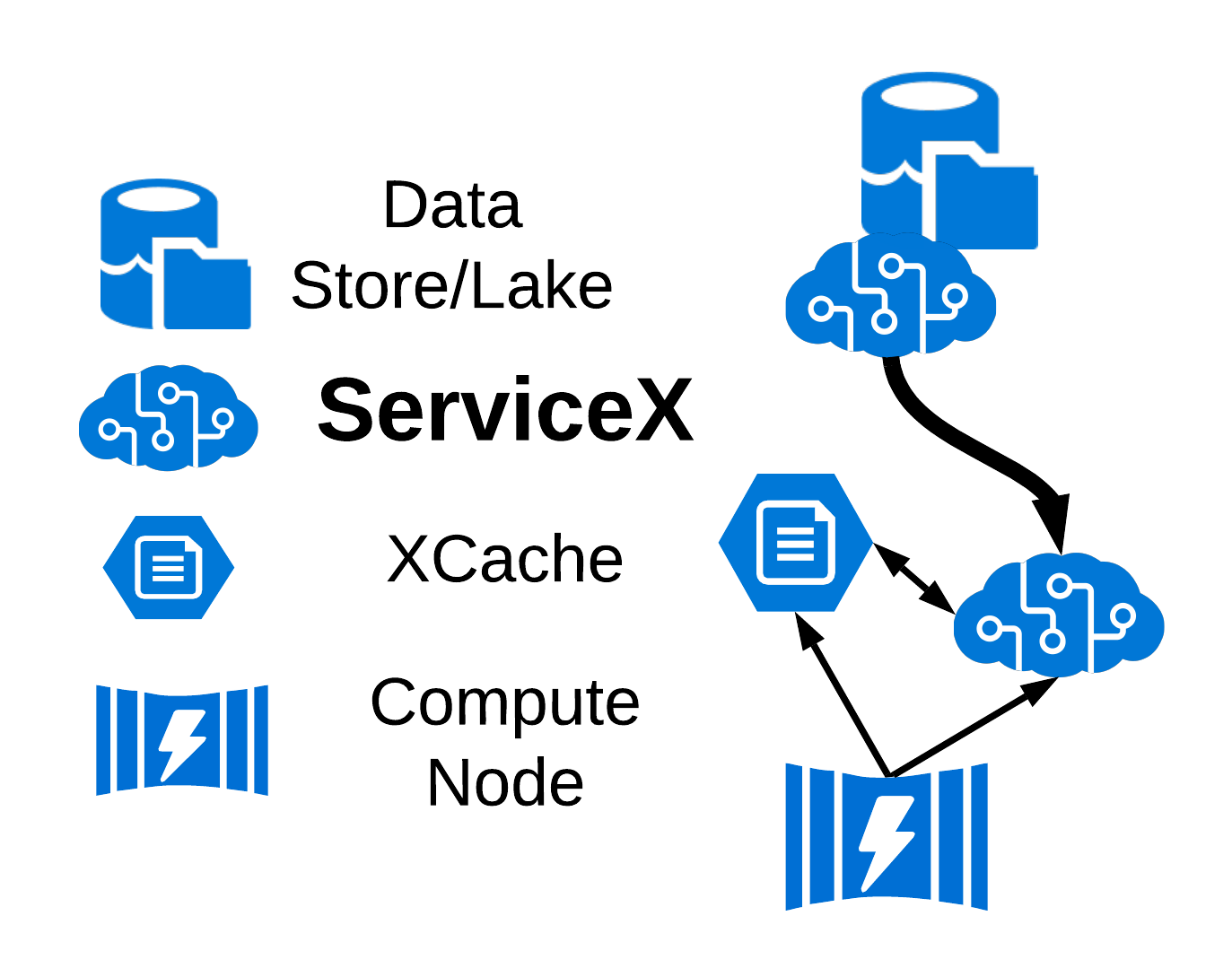}
\caption{ServiceX, a data delivery service from the IRIS-HEP DOMA area.}
\label{fig:idds}
\end{center}
\vspace{-.4in}
\end{wrapfigure}
%\end{figure}

Models of data organization, management and access for the extreme scales 
anticipated for the HL-LHC era are being developed
including ServiceX ~\cite{servicex}, a so-called ``intelligent data delivery service" developed by IRIS-HEP.
Its role in an overall delivery network is shown in Figure~\ref{fig:idds}.
It provides a flexible transformation layer between cold-storage formats (optimized for storage) and
those more suitable for scalable analysis platforms using modern processing frameworks and data structures, including
columnar arrays. It is composed of several individual containerized and autoscaling components:
input data finders, data transformers (with acceleration potentially provided by integrated device servers), 
queues, caches, monitoring, output services, etc.  Each component can be run at multiple sites,
and individually tuned for performance given the available resources and tasks being processed.
Specifically the deployed service components on SSL included:  Kafka brokering, RabbitMQ message bus, MinIO object, XCache 
data cache, and a Rucio (catalog) lookup service.  The deployment approach was to be
able to support data services embedded at various points in the 
distributed LHC cyberinfrastructure, deployed and operated by a central 
team of service developers.  Systems of orchestrated, containerized services 
were functionally tested and assessed for scalability and performance in
realistic configurations using leveraged resources from the participating 
institutions and the U.S. LHC computing facilities (in this case, the ATLAS Midwest Tier2 Center and the RIVER 
cluster~\cite{river} provided by the department of Computer Science at the University of Chicago).

\section{A Blueprint Process for an SSL}
A  `blueprint' workshop~\cite{as-ssl-bp} focused on requirements for supporting 
the Analysis Systems area to achieve IRIS-HEP year~2 R\&D deliverables and milestones helped drive
the SSL design. 
The meeting included talks from computer scientists (e.g. hardware acceleration for data delivery), 
industry partners (e.g. Google, Redhat) and resource providers at universities and HPC centers with whom IRIS-HEP is engaging for SSL resources. The \textbf{major goals} of the workshop included reviewing the status of the Analysis Systems (AS) milestones and deliverables to inform the needs for a collaborative development and testing platform; 
develop the SSL architecture and plans, using AS R\&D activities as specific examples; development of requirements on SSL to support the AS area, particularly the prototyping, benchmarking and scaling of AS deliverables toward production deployment; increase the visibility of SSL and AS beyond IRIS-HEP to facilitate partnerships with organizations that might provide software and computing resources toward these objectives; get informed on latest developments in open source technologies and methods important for the success of the SSL and AS R\&D areas of the Institute. \textbf{Key Outcomes} incuded identification of Kubernetes identified as a {\it common denominator} 
technology for the SSL, increasing our innovation capability through flexible infrastructure; plans for a multi-site SSL {\it substrate project} that will federate SSL contributions from multiple resource providers (institutes and public cloud), offering the AS area a flexible platform for service deployment at scales needed to test the viability of system designs; a vision for an SSL that serves as an innovation space for AS developers, and a testbed to prototype next generation infrastructure patterns for future HEP computing environments, including future LHC Tier2 centers.

\section{SSL Architectural Principles}
There are a number of desirable features that have been identified for the SSL.  These include: community access - open to all working on software infrastructure in HEP - which can be implemented with federation tools based on {\sf CI-Logon}, for example, providing a single sign-on capability using one's home institution credentials; a lightweight group (project) management system; 
infrastructure that is itself \textit{composable and reusable}; ability to accommodate and 
aggregate a diverse resource pool and user community; being flexible, agile, and dynamic; and ease of integration with public cloud resources when needed.  

It was clear that many of these features could be realized with 
a container-based service orchestration framework on 
dedicated resources, plus a {VC3}-like technology~\cite{vc3} to connect to HPC/HTC resources for batch scale-out. 
Open-source, cloud native technology, leveraging significant experience with the widely-adopted Kubernetes container orchestration software and associated ecosystem of container technologies, was the obvious choice for the SSL's base (`substrate') platform. 
This was not entirely clear before the blueprint workshop.

Regarding declarative and reproducible deployments, the goal is to have infrastructure built under the SSL to be easily reusable and deployable to other sites. In short, we require the SSL to provide resources that are \textit{discoverable}, \textit{flexible}, and \textit{nimble}. The declarative nature of Kubernetes is a good fit to the SSL requirments and gets us a long way to providing such a service. The SSL itself is \textit{not} intended to become a \textit{production center}. Rather, it should serve as an \textit{incubator} for projects which then \textit{graduate} to become full-fledged infrastructures that run on production resources. Services to build and manage artifacts -- tools that provide SSL to be scaled up and then back down -- are part of reducing cognitive load for developers and deployers.

\begin{wrapfigure}{r}{0.5\textwidth}
\begin{center}
\vspace{-.2in}
\includegraphics[width=0.5\textwidth]{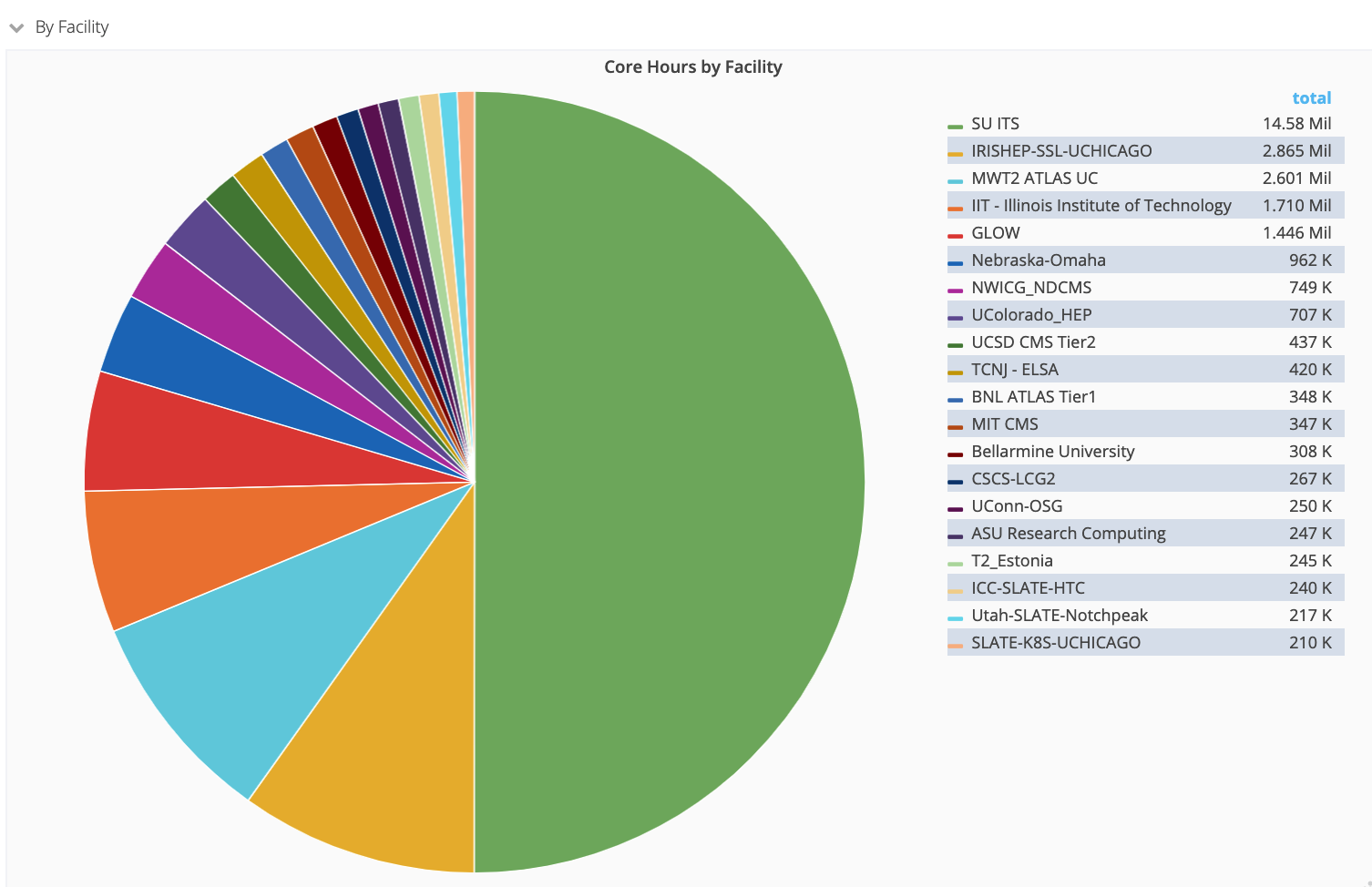}
\caption{Backfilling the IRISHEP-SSL-UCHICAGO cluster resource with high throughput jobs from the OSG virtual orgnization for a recent 90 day period in which 2.865M CPU-hours were delivered to a number of research groups.}
\label{fig:osg-hours}
\end{center}
\vspace{-.2in}
\end{wrapfigure}

Containerized services on the SSL follow the standard Kuberntes
model of being organized into ``Pods'', where multiple containers that need to
share some resources will be co-scheduled onto the same physical machine. Pods
will be organized into deployments, which will dictate policy for rolling out
pods and scaling parameters. Finally, ``Service'' objects will sit in front of
pods and provide externally-facing network access to user workloads.
Using annotations on services, users can preferentially
select, or require, workloads be run at a particular site. This may be a
desirable feature if, for example, a workload needs to be provisioned nearer
to storage external to the platform.  Registering an SSL cluster as a resource in SLATE 
allows for integration with existing computing efforts such as the Open Science Grid.
Other e-infrastructures are also possible; the OSG was chosen for proximity and 
familiarity with its service interfaces.

\begin{figure}[ht]
\begin{center}
\includegraphics[width=1.0\linewidth]{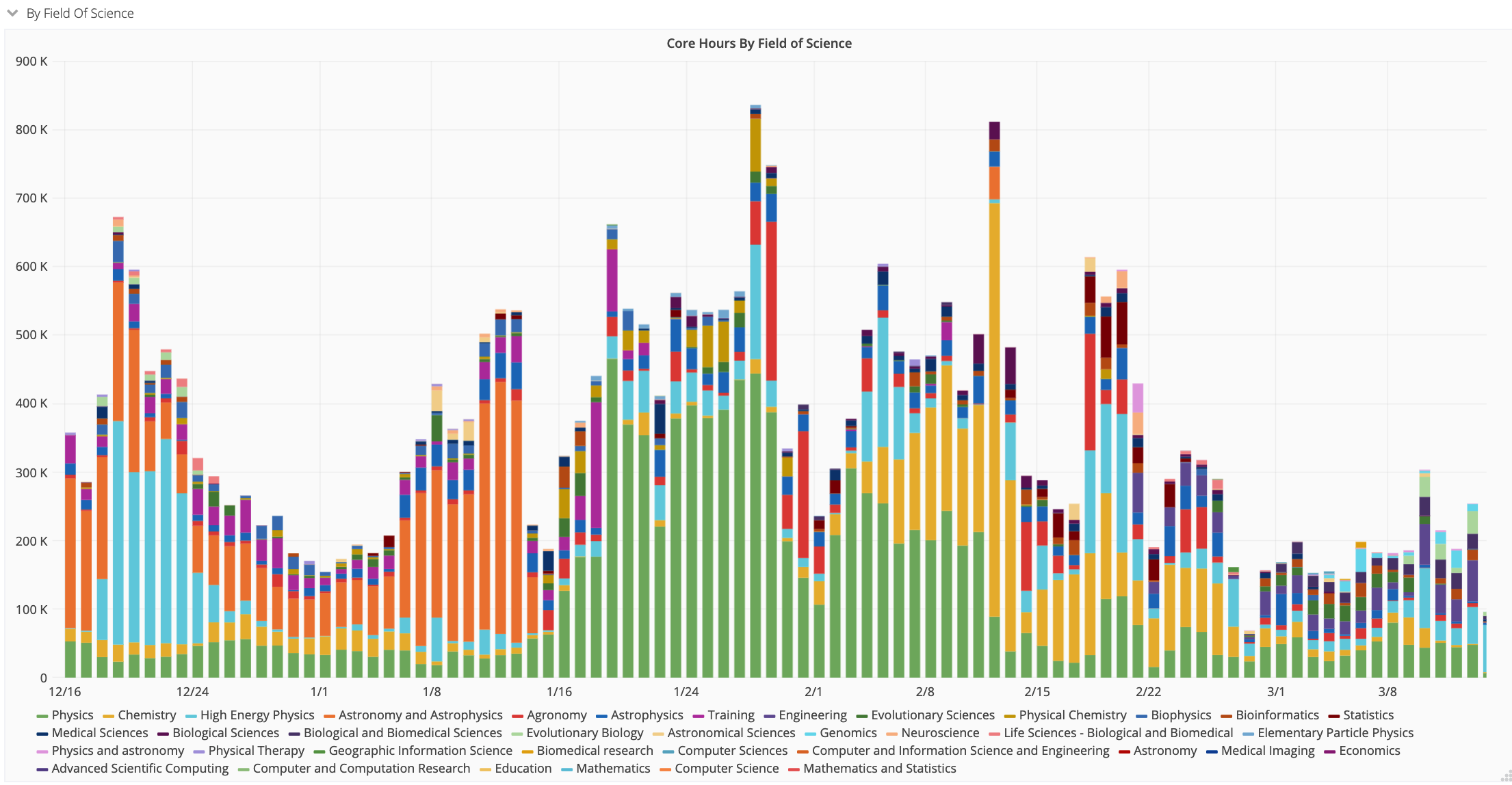}
\end{center}
\caption{Supported applications on the SSL by science domain when running in backfilling mode for open science.}
\label{fig:osg-fos}
\end{figure}

The SSL team initially used Google's Kubernetes Engine -- the Google Cloud Platform (GCP) -- to test ServiceX deployments.
This was pulled off GCP and onto in-house resources when they became available, running on Kubernates and integrated into the SSL.
Lightweight mechanisms are suggested for discovery of resources.  
The value of public display and reporting of science happening on contributed 
resources, to incentivize potential resource contributors, cannot be underestimated.  
For example, when not used for development and testing, 
SSL resources can be easily configured to be backfilled with workloads taken from the 
common Open Science Grid virtual organization user queue.  
Figure~\ref{fig:osg-hours} shows the (second leading) CPU-hours delivered by the SSL relative to other resource providing sites in the OSG, and
Figure~\ref{fig:osg-fos} shows where the contributions went according to science discipline. 
This demonstrates the flexibility to accommodate development and production in a single environment 
(even for diverse workloads) and thus efficiently using the
investment.  Kubernetes pods can be preempted by priority, such that the backfill jobs can be ejected 
when the system was needed for large scale testing of R\&D service and application deployments.

\section{SSL as Training Platform}

The SSL group has developed a versatile machine learning 
platform, capable of deployment across a number of Kubernetes resource targets.
It consists of several (containerized) services that can be individually customized, combined and 
deployed on any Kubernetes cluster.  This enables easy resource sharing and user authentication.
User-friendly interfaces and environments are convenient not only to researchers but also 
for lecture/workshop organization.
The platform provides private JupyterLab instances (with ATLAS and HEP community customized environments), 
Spark cluster deployments, and OpenAI environments. It will soon be 
extended with Tensorflow As a Service (TFaaS), Kubeflow, and a general AB testing support service.
The platform was extended to support CoDaS-HEP workshops~\cite{codashep},
with the backend configured to schedule notebooks to the GPU resources of the 
Pacific Research Platfrom (PRP)~\cite{prp} using the Kubernetes API.
The PRP provided 34 nodes, each with two Nvidia 1080 Ti processors, and were accessed by 55 students working 
through a 14 module minicourse in PyTorch, an open source machine learning framework. 

\begin{figure}[ht]
\begin{center}
\includegraphics[width=0.65\linewidth]{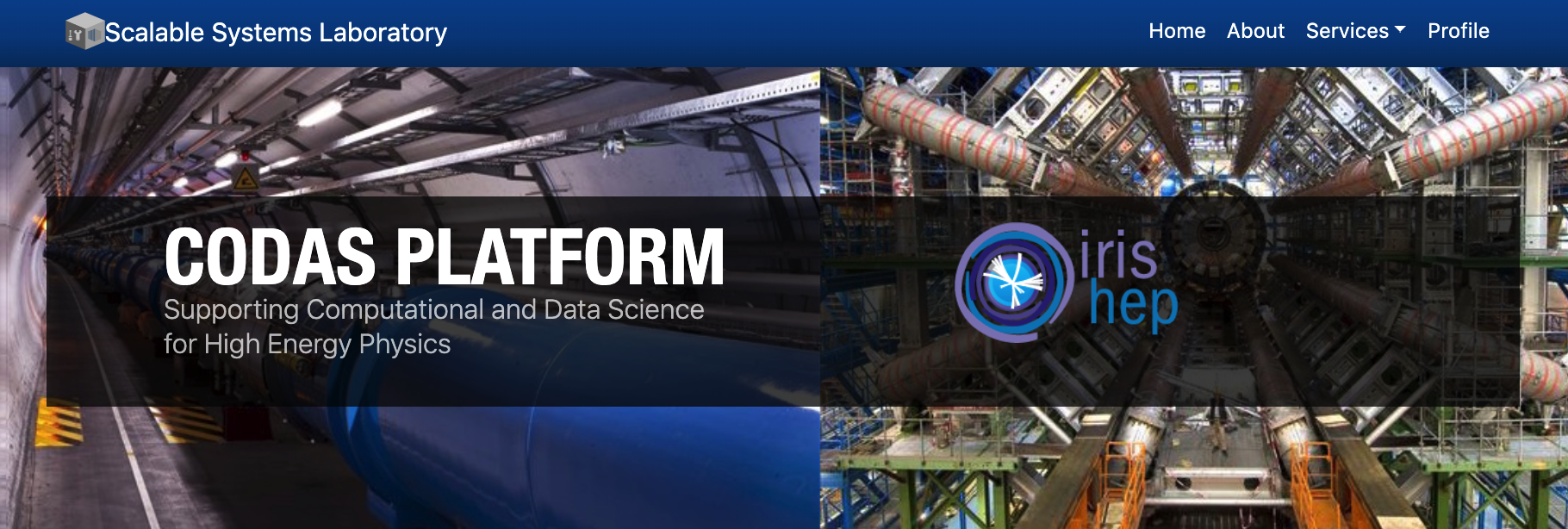}
\quad
\includegraphics[width=0.3\linewidth]{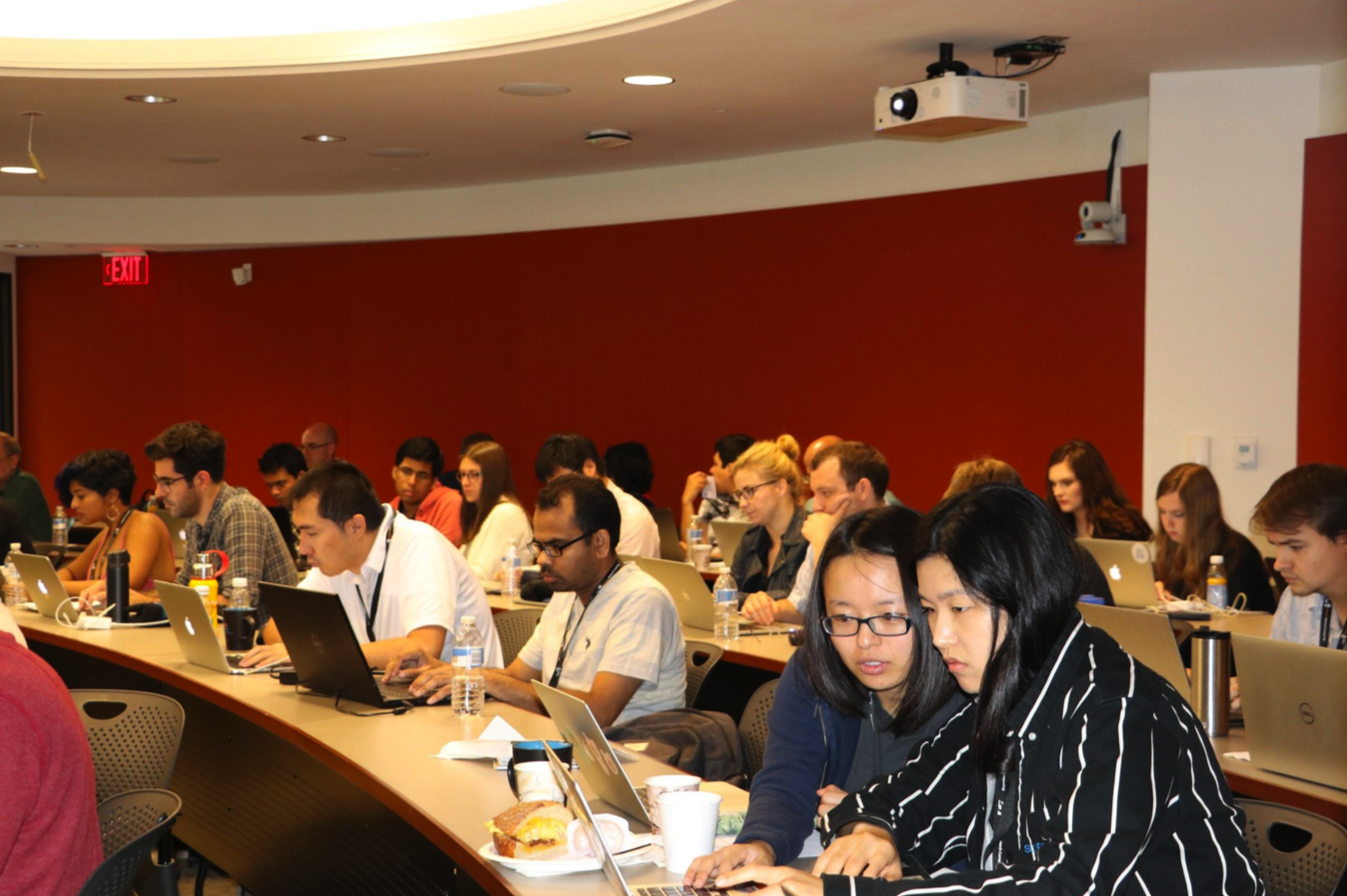}
\end{center}
\caption{Left: the web portal to a (notebook interface) machine learning platform which has been used for the Computational and 
Data Science for High Energy Physics (CoDaS-HEP) summer course at Princeton University, pictured at right.  
}
\label{fig:twoup}
\end{figure}

\section{Applications}

In Table~\ref{tab:apps} we list the number of ``applications'' (in most cases compositions of services) deployed to date on the SSL.  Each of these were containerzed with Docker and Helm by various groups from within the Institute and by LHC partners.   We note the deployment cycle used in each case: those under heavy development were repeatedly deployed and tested, while other more mature services could be operated over longer periods, in a quasi-production mode. 

\begin{table}[ht]
\begin{center}
\caption{Applications and services deployed on the IRIS-HEP SSL resource.}
\label{tab:apps}
\begin{tabulary}{1.00\textwidth}{|p{1.3cm}|p{2cm}|p{5.5cm}|p{1.75cm}|}
\hline
\textbf{Area} & \textbf{Application} & \textbf{Description}    &  \textbf{Deployment Cycles}  \\ 
\hline
 DOMA  &    ServiceX   &  Data transformation and delivery service for LHC analyses &  Development \& scale test \\ \hline
 DOMA  &  Skyhook   & Programmable storage for databases, scaling Postgres with Ceph object store  & Development \\ \hline
 AS  &   Parsl/FuncX   & Parallel programming in Python, serverless computing with supercomputers  & Development \\ \hline
 Comp. Sci.  &  Large Scale Systems   & Serverless computing with Kubernetes  & Development \\ \hline
 AS  &  REANA   & Reusable Analysis Service  & Production \\ \hline
 Training  &  CoDaS-HEP Platform   &  JupyterLab notebooks, access to GPU resources on the Pacific Research Platform for annual summer CoDaS-HEP training event & Production  \\ \hline
 OSG-LHC  &  SLATE Backfiller  &  Backfilling otherwise unused cycles on SSL with work from the OSG using the SLATE tools  &  Production \\ \hline
 ATLAS Exp. &    Frontier Analytics   &  Analyze and improve data access patterns for ATLAS Conditions Data &  Production \\ \hline
 Networks  &    perfSONAR Analytics   &  Network route visualization based on perfSONAR traces &  Production \\ \hline
\end{tabulary}
\end{center}
\vspace{-1.0em}
\end{table}

\section{Summary and Future}

\begin{figure}[th]
\centering
\includegraphics[width=\textwidth]{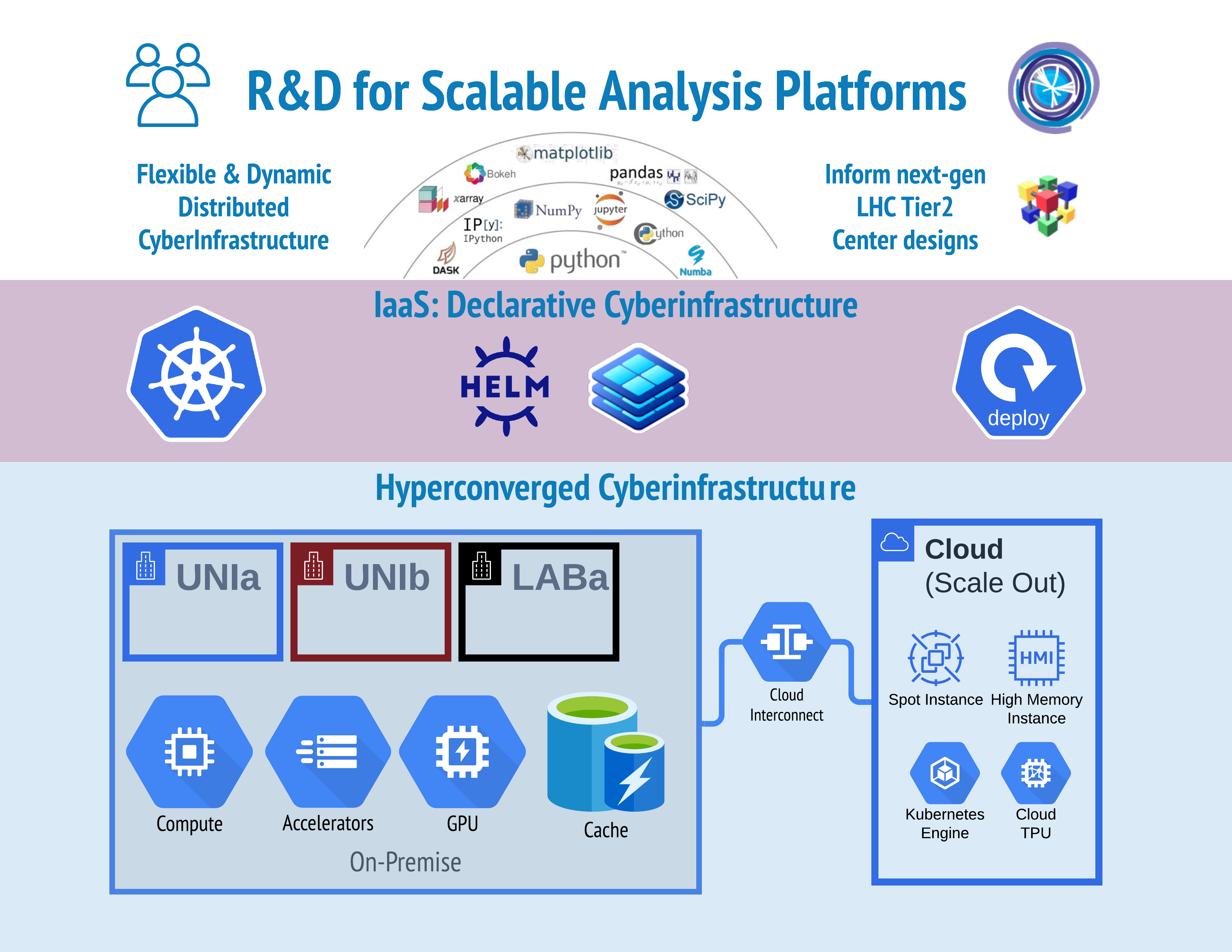}
\caption{The SSL will be used to prototype distributed CI deployment patterns, resulting in declarative, flexible and
scalable platforms.} 
\label{fig:deci}    
\vspace*{-15pt}
\end{figure}

In future we imagine SSL facilities formed from distributed cyberinfrastructure constructed with declarative principles, 
fully realizing
the distributed ``substrate'' model described above.  
A vision for this is illustrated schematically in Figure~\ref{fig:deci}.   
Software packages on the top, ``scalable platforms'' layer are deployed
on a declarative Infrastructure-as-a-Service (IaaS) platform which can span multiple 
computing resources (sites) and public cloud providers.
Through such a hyperconverged infrastructure, SSL will ``stretch'' a single
Kubernetes cluster across multiple sites, including public cloud for 
potential scale-out situations, giving users a single interface via
command-line client or RESTful API. Users will then be able to orchestrate
Docker containers from their own laptop. This obviates the need for users to
have a ``login node'', synchronize data to a remote machine, etc. 
Beyond the default Kubernetes installation, we plan to: add a number of additional
components to give users a more cloud-like experience, provide 
monitoring for operators, and augment the SSL with additional
capabilities to facilitate research. Experience gained will help inform the
shape of LHC Tier2 facilities in the HL-LHC era.

\section{Acknowledgements}
This work was supported by the National Science Foundation under Cooperative Agreement OAC-1836650 and by grant number OAC-1724821.
The CoDaS-HEP training platform was supported in part by NSF awards CNS-1730158, ACI-1540112, ACI-1541349, and OAC-1826967.

% \newpage

%
% BibTeX or Biber users please use (the style is already called in the class, ensure that the "woc.bst" style is in your local directory)
\bibliography{bibliography}

\end{document}